\documentclass[aps,pra,twocolumn]{revtex4-1}
\usepackage{epsfig,amssymb,amsmath,mathrsfs,color}
\usepackage[active]{srcltx}
\usepackage[hypertex,linkcolor=red]{hyperref}
\usepackage[matrix,frame,arrow]{xypic}
\usepackage[matrix,frame,arrow]{xy}
\usepackage{amsmath}

\newcommand{\qw}[1][-1]{\ar @{-} [0,#1]}

\newcommand{\gate}[1]{*{\xy *+<.6em>{#1};p\save+LU;+RU **\dir{-}\restore\save+RU;+RD **\dir{-}\restore\save+RD;+LD **\dir{-}\restore\POS+LD;+LU **\dir{-}\endxy} \qw}

\newcommand{\multigate}[2]{*+<1em,.9em>{\hphantom{#2}} \qw \POS[0,0].[#1,0];p !C *{#2},p \save+LU;+RU **\dir{-}\restore\save+RU;+RD **\dir{-}\restore\save+RD;+LD **\dir{-}\restore\save+LD;+LU **\dir{-}\restore}
\newcommand{\ghost}[1]{*+<1em,.9em>{\hphantom{#1}} \qw}

\newcommand{\Qcircuit}[1][0em]{\xymatrix @*[o] @*=<#1>}  
 \renewcommand{\Qcircuit}[1][0em]{\xymatrix @*=<#1>}

\newcommand{\pureghost}[1]{*+<1em,.9em>{\hphantom{#1}}}
\newcommand{\multiprepareC}[2]{*+<1em,.9em>{\hphantom{#2}}\save[0,0].[#1,0];p\save !C
  *{#2},p+RU+<0em,0em>;+LU+<+.8em,0em> **\dir{-}\restore\save +RD;+RU **\dir{-}\restore\save
  +RD;+LD+<.8em,0em> **\dir{-} \restore\save +LD+<0em,.8em>;+LU-<0em,.8em> **\dir{-} \restore \POS
  !UL*!UL{\cir<.9em>{u_r}};!DL*!DL{\cir<.9em>{l_u}}\restore}
\newcommand{\prepareC}[1]{*{\xy*+=+<.5em>{\vphantom{#1\rule{0em}{.1em}}}*\cir{l^r};p\save*!L{#1} \restore\save+UC;+UC+<.5em,0em>*!L{\hphantom{#1}}+R **\dir{-} \restore\save+DC;+DC+<.5em,0em>*!L{\hphantom{#1}}+R **\dir{-} \restore\POS+UC+<.5em,0em>*!L{\hphantom{#1}}+R;+DC+<.5em,0em>*!L{\hphantom{#1}}+R **\dir{-} \endxy}}
\newcommand{\poloFantasmaCn}[1]{{{}^{#1}_{\phantom{#1}}}}

 \def\map#1{\mathcal #1}
\def\d{\operatorname{d}}\def\<{\langle}\def\>{\rangle}
\def\Tr{\operatorname{Tr}}\def\:{\hbox{\bf
    :}}

\def\spc#1{\mathscr{#1}}

\begin{document}
\title{Perfect discrimination of no-signalling channels  via quantum superposition of causal structures} 

\author{Giulio Chiribella} \affiliation{ Center for Quantum Information, Institute for Interdisciplinary Information Sciences, Tsinghua University, Beijing 100084, China   } 
  
 \date{ \today}
\begin{abstract} 
A no-signalling channel transforming quantum systems in Alice's and Bob's local laboratories is compatible with two different  causal structures:  ($A\preceq B$) Alice's output causally precedes  Bob's input and  ($B\preceq A$) Bob's output causally precedes Alice's input.     Here I prove that two no-signalling channels that are not perfectly distinguishable in any ordinary quantum circuit can become perfectly distinguishable  through the quantum superposition of circuits with different causal structures.

\end{abstract}
\maketitle
Distinguishing between two objects is one of the most fundamental tasks in information theory.  In  Quantum Information, an instance of the problem is the discrimination between two quantum channels \cite{acin,lopresti,max1,max2,duanseq,duanloc,memorydisc,illupol,illugauss,shapiro,duanqo,harrow}:  In this  scenario one has access to a black box implementing a transformation of quantum systems, which is promised to be either $\map C_0$ or $\map C_1$, and the goal is to identify such a transformation with maximum probability of success using a given  number of queries to the black box.

Many surprising features of quantum channel discrimination have been discovered so far.   For example, two unitary channels that are not perfectly distinguishable with a single query become perfectly distinguishable when a finite number of queries  is allowed \cite{acin,lopresti}. 
Other remarkable phenomena arise when the two channels $\map C_0$ and $\map C_1$ have a bipartite structure, 
as in the following diagram 
\begin{align*}
  \Qcircuit @C=1em @R=.7em @! R {&\qw \poloFantasmaCn{A} &\multigate{1} { \map C_i}   & \qw \poloFantasmaCn{A'} &\qw\\
    &\qw \poloFantasmaCn{B}& \ghost{\map C_i} & \qw \poloFantasmaCn{B'}  &\qw }\qquad i =  0,1 
\end{align*}  
which represents a quantum channel (i.e. a completely positive trace preserving map) sending quantum states on the Hilbert space $A \otimes B$ to quantum states on the Hilbert space $A' \otimes B'$. We can imagine that the channels $\map C_0$ and $\map C_1$  transform quantum states provided by two users, Alice and Bob.  If the state of Alice's output  $A'$ does not depend on the state of  Bob's input  $B$, then we say that $\map C_0$ and $\map C_1$ are  \emph{no-signalling from Bob to Alice} (\emph{$B$-no-signalling}, for short).        Eggeling, Schlingemann, and Werner \cite{esw} showed that every B-no-signalling channel $\map C$   can be realized as the concatenation of a channel $\map A$  on Alice's side followed by a channel $\map B$ on Bob's side, with some information transferred from Alice to Bob via a quantum memory $M$, as in the diagram:     
\begin{align}\label{esw}
 \begin{aligned}
  \Qcircuit @C=1em @R=.7em @! R {&\qw \poloFantasmaCn{A} &\multigate{1} { \map C}   & \qw \poloFantasmaCn{A'} &\qw\\
    &\qw \poloFantasmaCn{B}& \ghost{\map C} & \qw \poloFantasmaCn{B'}  &\qw }
    \end{aligned}  
    = \begin{aligned}
  \Qcircuit @C=1em @R=.7em @! R {
  &\qw \poloFantasmaCn{A} &\multigate{1} { \map A}   & \qw \poloFantasmaCn{A'} &\qw  &    &  \qw \poloFantasmaCn{B}  &  \multigate{1}{\map B}  & \qw \poloFantasmaCn{B'} & \qw  \\
   &    &   \pureghost{\map A}  &  \qw \poloFantasmaCn{M}    &  \qw & \qw &\qw  & \ghost{\map B}  &   & }
    \end{aligned} 
    \end{align}  
 In other words, if a channel does not signal from Bob to Alice, then it is compatible with a causal structure where Alice's output  precedes Bob's input, denoted by $A \preceq B$:  In this structure the black box provides Alice's output before Bob supplies his input, as illustrated in the r.h.s. of Eq. (\ref{esw}).     To discriminate between two B-no-signalling channels  we can then use a \emph{sequential strategy}  \cite{memorydisc},  where the channel $\map C_i$ (either with $i= 0$  or with $i=1)$ is inserted in a quantum circuit with  causal structure $A \preceq B$, thus producing the output state $\rho^{seq}_i$ given by 
 \begin{equation}\label{seq}
 \begin{aligned}
\Qcircuit @C=1em @R=.7em @! R {
\multiprepareC{1}{\rho^{seq}_i}  & \qw \poloFantasmaCn{R'}  & \qw \\
\pureghost{\rho^{seq}_i}  &    \qw \poloFantasmaCn{B'} & \qw }
\end{aligned}   :=
  \begin{aligned}\Qcircuit @C=1em @R=.7em @! R {
   \multiprepareC{1}  {\Psi}  &  \qw \poloFantasmaCn{R}  &  \qw & \qw  &  \multigate{1}{\map W}  &  \qw \poloFantasmaCn{R'}  & \qw  & \qw &\qw \\
   \pureghost{\Psi}  &  \qw \poloFantasmaCn{A}    &  \multigate{1}{\map A_i}  &  \qw \poloFantasmaCn{A'}  &  \ghost{\map W}  &  \qw \poloFantasmaCn{B}  &  \multigate{1}{\map B_i}  &  \qw \poloFantasmaCn{B'}  & \qw \\
   &&  \pureghost{\map A_i}  &  \qw \poloFantasmaCn{M_i}   & \qw &  \qw &  \ghost{\map B_i} & &    }
\end{aligned}  
\end{equation}     
Here $R$ and $R'$ are suitable ancilary systems, $M_i$ is the quantum memory needed for the realization of channel $\map C_i$, $\Psi$ is a pure state on $R \otimes A$ and $\map W (\rho)= W \rho W^\dag $, $W^\dag W  =  I_{R \otimes A'}$ is an isometry sending states on $R \otimes A'$ to states  on $R'\otimes B$ \cite{purificaz}. 

Ref. \cite{memorydisc} showed that sequential strategies offer an advantage over \emph{parallel strategies}, where the channel $\map C_i$ is applied  on one side of an entangled input state $\Psi \in  A \otimes B \otimes R$, producing the output state $\rho_i^{par}$ given by
\begin{align}\label{par}
 \begin{aligned}
\Qcircuit @C=1em @R=.7em @! R {
\multiprepareC{1}{\rho^{par}_i}  & \qw \poloFantasmaCn{R'}  & \qw \\
\pureghost{\rho^{par}_i}  &    \qw \poloFantasmaCn{B'} & \qw }
\end{aligned}   :=
\begin{aligned}
\Qcircuit @C=1em @R=.7em @! R { 
\multiprepareC{2}{\Psi} & \qw\poloFantasmaCn{A}  &  \multigate{1}{\map C_i}  & \qw \poloFantasmaCn{A'} &  \qw  \\
\pureghost{\Psi}  &  \qw \poloFantasmaCn{B}  & \ghost{\map C_i}  & \qw  \poloFantasmaCn{B'} &  \qw  \\
\pureghost{\Psi}  & \qw \poloFantasmaCn{R}  & \qw & \qw & \qw}  
\end{aligned} 
  \end{align}
 In particular, Ref.  \cite{memorydisc} exhibited two B-no-signalling channels that can be perfectly distinguished by a sequential strategy, whereas every parallel strategy has a non-zero probability of error.  Later,  Harrow \emph{et al} \cite{harrow} demonstrated  the same phenomenon in the absence of a quantum memory,  i.e. for two channels $\map C_0$ and $\map C_1$ of the product form $\map C_i  =  \map A_i \otimes \map B_i$, with $\map A_i $  ($\map B_i$) transforming states on $A$ ($B$) into states on $A'$ ($B'$).  Channels of this form are a particular example of \emph{no-signalling channels}  \cite{piani,facco}, namely,  channels that are both B-no-signalling and A-no-signalling  \cite{nosig}.

 In principle, no-signalling channels can be used in two different causal structures:  $A \preceq B$ (the black box processes Alice's input first and  Bob's input later) and $B \preceq A$ (the black box processes Bob's input first and  Alice's input later).  Usually, when there are two possible alternatives, in quantum theory one can conceive a superposition of them.  Can we apply this idea also to the choice of causal order?   Recently,  Ref. \cite{qswitch} introduced the notion of \emph{quantum superposition of causal structures}, arguing  that this new primitive could be achieved in a quantum network where the connections among devices are controlled by the quantum state of a control qubit.  A no-signalling channel inserted in such a network would be in a quantum superposition of being used in a circuit with causal structure $A\preceq B$ and of being used in a circuit with  causal structure $B \preceq A$.   
 Such a network can be thought as a toy model for a quantum gravity scenario where the causal structure is not defined a priori, a scenario originally considered  by Hardy \cite{hardy}, who posed the question whether indefinite causal structure can be used as a computational resource.



This paper  gives a positive answer to Hardy's question, showing that the superposition of causal structures enables completely new schemes  for quantum channel discrimination.    The advantage of such schemes  is demonstrated by exhibiting a concrete example of two no-signalling channels that  cannot be perfectly distinguished by any sequential strategy using a single query to the black box, but become perfectly distinguishable  through a quantum superposition of sequential strategies with different causal structures \cite{qswitch}. The example involves two-qubit channels $\map C_0$ and $\map C_1$, with $\map C_0$  consisting of two von Neumann measurements on the same random basis, and $\map C_1$ consisting of two rotations of $\pi$ around a random pair of orthogonal axes in the Bloch sphere.  More generally, the superposition of causal structures presented here allows for a single-query, zero-error discrimination between an arbitrary pair of qubit channels with commuting Kraus operators and an arbitrary pair of qubit channels with anticommuting Kraus operators.  These  results  contribute to the exploration of a new research avenue that aims at demonstrating new physical phenomena and power-ups to information processing  arising from the application of quantum theory in the lack of a definite causal structure   \cite{hardy,qswitch,costa}.

Before presenting the result, let us make precise  what we mean by quantum superposition of
causal structures.  We can start from the simplest case, where the channels $\map C_0$ and $\map C_1$ are of the product form $\map C_i =  \map A_i \otimes \map B_i$, $i=0,1$.   First, let us  write down the Kraus forms $\map A_i (\rho) =
\sum_k A_{ik} \rho A_{ik}^\dag$ and $\map B_i (\rho)= \sum_l B_{il}
\rho B_{il}^\dag$. For each value of $k$ and $l$, a  sequential circuit with causal structure $A \preceq B$   (like the circuit in Eq.
(\ref{seq})) yields the (unnormalized) pure state
\begin{equation}\label{out1}
  |\Psi_{ikl} \>  =     (B_{il} \otimes I_{R'}) W (A_{ik} \otimes I_R) |\Psi\>.
\end{equation} 
whereas a sequential circuit  with causal structure $B \preceq A$  yields 
\begin{equation}\label{out2}
  |\widetilde \Psi_{ikl} \>  =     (A_{ik}  \otimes I_{ \widetilde R'}) \widetilde W (B_{il} \otimes I_{\widetilde R}) |\widetilde \Psi\>,
\end{equation}
where $\widetilde R$ and $\widetilde R'$ are suitable ancillary systems,  $\widetilde \Psi \in  B \otimes \widetilde R$ is a pure state and $\widetilde W$ is an isometry from $B' \otimes \widetilde R$ to $A \otimes \widetilde R'$. Note that, by suitably choosing the ancillary systems $R, R', \widetilde R, \widetilde R'$ we can assume without loss of generality that the input and output systems of the two circuits are the same, i.e.    $A \otimes R \simeq B \otimes \widetilde R $ and  $B \otimes R' \simeq A \otimes \widetilde R' $.  Suppose now that we have at disposal a coherent mechanism that chooses
the first circuit when the state of a control qubit is $|0\>$, and the
second when the state is $|1\>$. 
Such a mechanism could be implemented in a quantum network where the connections among devices are not pre-determined, but instead can be controlled by the state of some quantum system, as in the \emph{quantum switch} of Ref. \cite{qswitch}.
If the control qubit is prepared in the state $|\psi\> =  \alpha |0\> +\beta |1\>$, then
the output of the network is $ \alpha |\Phi_{ikl} \> |0\> + \beta |\Phi'_{ikl}
\>|1\> $. Taking the corresponding density matrix and
summing over all possible Kraus elements we then get the output state
\begin{align}
 \nonumber \rho^{sup}_{i}: =  &  |\alpha|^2  (\map B_i \otimes \map  I_{R'} )\map W (\map A_i \otimes \map I_R) (|\Psi\>\<\Psi|) \otimes |0\>\<0|  \\ 
  \nonumber & +| \beta|^2  (\map A_i \otimes \map I_{R'} )  \widetilde {\map  W}(\map B_i
  \otimes \map I_{R}) (|\widetilde {\Psi} \>\<\widetilde {\Psi}|) \otimes |1\>\<1| \\
\nonumber &+\alpha\beta^*   \sum_{k,l}  |\Psi_{ikl} \>\< \widetilde { \Psi}_{ikl} | \otimes |0\>\<1| \\
 &  + \alpha^* \beta \sum_{k,l}  |\widetilde {\Psi}_{ikl} \>\< \Psi_{ikl} | \otimes |1\>\<0|  , \label{superp}
\end{align}
with $\widetilde {\map W}  (\rho)  =  \widetilde W \rho \widetilde W^\dag$.   The first two terms in Eq. (\ref{superp}) are the classical ones, corresponding
to the random choice of two possible circuits with causal structures $A \preceq B$ and $B \preceq A$.   The off-diagonal terms have no classical interpretation:  they represent the
quantum interference between the two different causal structures. 
Note that the state in Eq. (\ref{superp}) does not depend on the
particular Kraus representation chosen for the channels $\map A_i$ and 
$\map B_i$: had we chosen another Kraus representation, after summation
we would have obtained the same result.   Eq. (\ref{superp}) can be  extended by linearity to the case of generic no-signalling channels, by writing the Kraus form $\map C_i  (\rho) =  \sum_m C_{im}  \rho  C_{im}^{\dag}$ and expanding the Kraus operators as $C_{i m}  =  \sum_{k}  A_{imk}  \otimes B_{imk}$.  

The availability of a  network implementing the quantum  superposition of causal structures can be interpreted as a new information-theoretic primitive that takes as input a query to a generic no-signalling channel $\map C_i$ and produces as output one query to the channel defined by $\map C_i^{sup}  (\rho)  := \rho_i^{sup}$, where $\rho_i^{sup}$ is the state defined in Eq. (\ref{superp}) and $\rho$ is the projector on the state $\alpha |\Psi \> |0\> +\beta  |\tilde \Psi
\>|1\> $. We will now show that having access to this primitive can reduce by a factor two the number of queries needed for the discrimination of a pair of no-signalling channels. 
In our example, all systems are qubits:  $A  \simeq A'  \simeq B  \simeq B'  \simeq  \mathbb C^2$.  Channel $\map C_0$  consists of two von Neumann measurements on the same  random basis
\begin{align}
\map C_0   &: =  \int \d U  ~  \map M^{(A)}_U  \otimes \map M^{(B)}_U,
\end{align}
where $\d U$ is the normalized Haar measure on $SU (2)$ and $ \map M_{U} $ is the single-qubit channel given by $\map M_U (\rho) :  =  \<  0| U^\dag  \rho   U |0\> ~  U|0\>\<0|U^\dag +  \<  1| U^\dag  \rho   U |1\> ~  U|1\>\<1|U^\dag $, $\{|0\>,|1\>\}$ being the computational basis.  Channel $\map C_1$ consists of two rotations of $\pi$ around a pair of random orthogonal axes in the Bloch sphere:  
\begin{align}
\map C_1  &: =  \int \d V   ~  \map X^{(A)}_V  \otimes \map Y^{(B)}_V,  
\end{align} 
where $\map X_V (\rho)  :  =  (V X V^\dag)  \rho  (VXV^\dag)$ and  $\map Y_V (\rho)  :  =  (V Y V^\dag)  \rho  (VYV^\dag)$, $X$ and $Y$ being the Pauli matrices representing rotations of $\pi$ around the $x$ and $y$ axes, respectively.  

Suppose that an experimenter has access to the bipartite black box and is asked to discriminate between  $\map C_0$ or $\map C_1$ using a single query.   The discrimination between $\map C_0$ and $\map C_1$ is equivalent to the discrimination between two product channels $\map C_{0, U}  :  =  \map A_{0,U}  \otimes \map B_{0, U}$  and $\map C_{1, V}  :  =  \map A_{1,V}  \otimes \map B_{1, V}$,   with $\map A_{0,U}  \equiv  \map B_{0,U} :=  \map M_U $  and   $\map A_{1,V}  :  =  \map X_V$ and $  \map B_{1,V} :=  \map Y_V $, where the unitaries $U$ and $V$ are completely unknown.      
To achieve perfect discrimination between $\map C_{0,U}$ and $\map C_{1,V}$ one can take a quantum superposition of the following two circuits:
\begin{equation}\label{duecircuiti}
  \begin{aligned} a) \Qcircuit @C=1em @R=.7em @! R {
      & \prepareC{\varphi} &  \gate{\map A_{i,U_i}} & \gate{\map B_{i, U_i}} &\qw }
\end{aligned} , ~
\begin{aligned} b) \Qcircuit @C=1em @R=.7em @! R {
    & \prepareC{\varphi} &  \gate{\map B_{i, U_i}} & \gate{\map A_{i, U_i}} & \qw } 
\end{aligned} 
\end{equation}     
where $\varphi$ is a fixed pure state and $U_0  :=  U$ and $U_1  : = V$.   The key idea  is that the Kraus operators of $\map C_0$ and $\map C_1$ behave very differently when we switch the ordering from $A \preceq B$ to $B \preceq A$: the Kraus operators of $\map A_{0,U} $ and $\map B_{0,U} $ commute for every $U$ (they are projectors on the same basis vectors),  whereas the Kraus operators of $\map A_{1,V}$ and  $\map B_{1,V} $  anticommute for every $V$. The difference between commutation and anticommutation cannot be detected by any ordinary circuit using a single query to the black boxes, but   becomes visible in the interference terms when we superpose the two circuits a) and b) with amplitudes $\alpha = \beta = \frac 1 {\sqrt 2}$:  Using Eqs.  (\ref{out1}), (\ref{out2}), and (\ref{superp})  with  $R \simeq R' \simeq \widetilde R  \simeq \widetilde R'  \simeq \mathbb C$, $\Psi  = \widetilde \Psi  = \varphi$, and $W =  \widetilde W  =  I$, we
obtain the output states
\begin{align*}
\rho^{sup}_0 &=  \map M_U  (|\varphi\>\<\varphi|)   \otimes |+\>\<+|    \\
\rho^{sup}_1  &=  \map Z_V ( |\varphi\>\< \varphi|)  \otimes |-\>\<-|  ~, 
\end{align*}
where $\map Z_V$ is the unitary channel $\map Z_V (\rho): =    (VZV^\dag) \rho  (VZV^\dag)$, $Z$ being the Pauli matrix $Z: =  -i XY$, and $|\pm\>  :  =  (|0\>  \pm  |1\>)/\sqrt 2$.  
By measuring the control qubit on the basis $|+\>,|-\>$ the experimenter can
perfectly distinguish between $\map C_0$  and $\map C_1$, \emph{no matter what the unknown unitaries $U$ and $V$ are}.   More generally, the above scheme allows one  to distinguish an arbitrary pair  $(\map A_0,\map B_0)$ of channels with  commuting Kraus operators $A_{0i} B_{0j}  =  B_{0j}  A_{0i}~  \forall i,j$ from an arbitrary pair $(\map A_1, \map B_1)$ of channels with anticommuting Kraus operators  $A_{0i} B_{0j}  =-  B_{0j}  A_{0i}  ~\forall i,j$.

We now show that no quantum circuit with fixed causal structure can perfectly distinguish between $\map C_0$ and $\map C_1$ with a single query.  The proof requires the formalism of \emph{quantum combs} \cite{combprl,combpra}, which describes the most general sequential strategies.  
This formalism makes extensive use of the Choi isomorphism \cite{choi} between a channel $\map C$ transforming states on $\spc H$ and the positive operator  $C$  on  $\spc H \otimes \spc H$  defined by $C:=  (\map C \otimes \map I)  (|I\>\!\>  \<\!\<  I|)$,  where $\map I$ is the identity map  and $|I\>\!\>$ is the  maximally entangled vector $|I\>\!\>:  = \sum_{n}  |n\>|n\>  \in  \spc H \otimes \spc H$, $\{|n\>\} $ being a fixed orthonormal basis for $\spc H$. In general, we will  use the ``double ket" notation $|\Psi\>\!\> :  =  (\Psi \otimes I)  |I\>\! \>$, where $\Psi$ is any operator on $\spc H$. 
Defining $\spc H_1  :=  A$,  $\spc H_2  :  = A'$, $\spc H_3:  = B$, $\spc H_4=  B'$,  the Choi operator of the channel $\map C_i$, $i= 0,1$  is the operator on  $\spc H_4 \otimes\spc H_3 \otimes\spc H_2 \otimes\spc H_1$ given by
\begin{align}\label{avechannel}
C_i   : =  \int \d U  ~  (\map U  \otimes \map U^*  \otimes \map U \otimes \map U^*) ( \Lambda_i )  
\end{align}
where 
$\Lambda_0  =   \sum_{m,n = 0,1}    |m\>\<m|  \otimes   |m\>\<m|  \otimes  |n\>\<n|  \otimes  |n\>\<n| $,   
$\Lambda_1= |Y\>\!\>  \<\!\<  Y|  \otimes   |X\>\!\> \<\!\<  X|$ ,
 and  $\map U$  ($\map U^*$) is the unitary channel defined by $\map U (\rho)  :  =U \rho U^\dag$  ($\map U^* (\rho) := U^*  \rho U^T$), $U^*$ ($U^T$) denoting the complex conjugate (the transpose) of the matrix $U$. 

Let us consider discrimination strategies with causal structure $A \preceq B$.   The discrimination is represented by a binary \emph{quantum tester} $\{T_0, T_1\}$, consisting of two positive operators $T_0$ and $T_1$ on $\spc H_4 \otimes\spc H_3 \otimes\spc H_2 \otimes\spc H_1$ that give the probabilities of the measurement outcomes according to the generalized Born rule $p(i|  \map C_j )  =  \Tr [T_i  C_j]$  \cite{memorydisc}.  The normalization of the tester is given by the condition  $T_0  +  T_1  =  I_4 \otimes \Xi$, where 
 $\Xi$ is a positive operator on  $\spc H_3 \otimes \spc H_2\otimes \spc H_1$ satisfying  the relation  
\begin{align}\label{normxi}
\Tr_3 [\Xi]  =   I_2 \otimes \rho,  \qquad \Tr[\rho]  =1,   
\end{align}
$\Tr_3$ denoting the partial trace over $\spc H_3$ and $\rho$ being a quantum state on $\spc H_1$ (see   Ref. \cite{memorydisc} for more details).  
Now, it is clear from Eq. (\ref{avechannel}) that the outcome probabilities are not affected if we replace each $T_i$, $i=0,1$with its average   $T'_i   : =  \int \d U  ~  (\map U  \otimes \map U^*  \otimes \map U \otimes \map U^*) ( T_i )$.   Since the average commutes with all the unitaries $U \otimes U^*  \otimes U \otimes U^*$, we can assume without loss of generality the commutation relation
\begin{align}\label{xicomm}
[\Xi,  U^*\otimes U \otimes U^*]  = 0 \qquad\forall U  \in  SU (2). 
\end{align}

From Ref. \cite{memorydisc}, we know that distinguishing between the two channels $  \map C_0$ and $\map C_1$  with the tester $\{T_0, T_1\}$   is equivalent to distinguishing between the two states $\Gamma_0$ and $\Gamma_1$ given by  
$\Gamma_i  : =    \left(I_4 \otimes \Xi^{\frac 12}   \right)   C_i   \left(  I_4 \otimes \Xi^{\frac 12}   \right) ,  i=  0,1$.
Hence,   $\map C_0$ and $\map C_1$ are perfectly distinguishable if and only if  $\Gamma_0$ and $\Gamma_1$ have orthogonal support, that is, 
$\Tr[\Gamma_0  \Gamma_1]  =  0 $.  Note that we have    
$\Tr[\Gamma_0  \Gamma_1]  =  \Tr \left[   \widetilde C_0 \left(I_4 \otimes \widetilde \Xi    \right)   \widetilde C_1   \left(  I_4 \otimes \widetilde \Xi   \right)  \right]$,
having defined $\widetilde C_i  :  = (I \otimes  Y  \otimes I \otimes  Y )  C_i  (I \otimes  Y \otimes I \otimes  Y ) $ for  $i=1,2$, and   $\widetilde \Xi  :=  (  Y \otimes I \otimes  Y)  \Xi  (  Y \otimes I \otimes Y) $.  

  We now prove that the condition $\Tr [\Gamma_0 \Gamma_1]  =0$ cannot be satisfied. First note that, by definition, $\widetilde \Xi$ must satisfy Eq. (\ref{normxi}) for some density matrix $\rho$.  Moreover,  from from Eq. (\ref{xicomm}) and from the relation $U^*  =  Y U Y, \forall U \in SU (2)$ it follows that $\widetilde \Xi$  must satisfy the commutation relation $[\widetilde \Xi  , U \otimes U \otimes U]  = 0$, $\forall U \in SU(2)$.       Hence, by the Schur lemmas  $\widetilde \Xi$ must be a combination of projectors onto the irreducible subspaces of $U\otimes U \otimes U$. The latter  are easily obtained by coupling the three angular momenta:  we have the subspace $\spc L_{\frac 32  }$ corresponding to $j=  \frac 32$  and  two subspaces $\spc L_{\frac 12}^{(1)}$ and  $ \spc L_{\frac 12}^{(0)}$  corresponding to $j=\frac 12$, which are spanned by the vectors $\{\Phi^{(1)}_{0}, \Phi^{(1)}_{1}  \}$ and  $\{\Phi^{(0)}_{0}, \Phi^{(0)}_{1}  \}$, respectively: 
\begin{align*}
|\Phi^{(1)}_{0}\>  &: =  \frac 1 {\sqrt {6}}  \left(   |0\>_{1}  |Y\>\!\>_{2,3}  +    |Y\>\!\>_{1,3}    |0\>_{2} \right)\\
|\Phi^{(1)}_{1}\>  &: =  \frac 1 {\sqrt {6}}  \left(   |1\>_{1}  |Y\>\!\>_{2,3}  +|Y \>\!\>_{1,3}     |1\>_{2}   \right)\\
|\Phi^{(0)}_{0}\>  &: =  \frac{  |Y\>\!\>_{1,2}  |0\>_3}{\sqrt 2}\\ 
 |\Phi^{(0)}_{1}\>  &: =  \frac{|Y\>\!\> _{1,2}   |1\>_3}{\sqrt 2}
\end{align*}

The most general expression for a positive operator $\widetilde \Xi$ commuting with $U \otimes U \otimes U$ is then
\begin{align}\label{xi}
\widetilde \Xi  =   a     P_{\frac 32 }     +   \sum_{m,n=0, 1}   b_{mn}   ~  T_{\frac 12}^{mn} , 
\end{align}
where $a\ge0$, $P_{\frac 32}$ is the projector onto $\spc L_{\frac 32}$,  $(b_{mn})$ is a positive two-by-two matrix, and 
$T_{\frac 12}^{ mn} :  =  \sum_{k=0,1}   |\Phi_{k}^{(m)} \>\< \Phi_{k}^{(n)}|$.  
Let us analyze now the normalization (\ref{normxi}).  First, note that the state $\rho$ in Eq. (\ref{normxi})  must be the invariant state $\rho  = I/2$  due to the symmetry $[\widetilde \Xi, U \otimes U \otimes U]=0$, which implies $[\rho, U]  =  0$.   
On the other hand, taking the partial trace we obtain $
\Tr_3[ P_{\frac 32} ]  =  \frac 43  P_{1} ,
\Tr_3[T_{\frac 12}^{ 11 }]  = \frac  23  P_1,
\Tr_3[T_{\frac 12}^{ 00}]  =   2  P_0  $, and 
$\Tr_3[T_{\frac 12}^{ 01}]  = \Tr[T_{\frac 12}^{ 10}]  =  0$.
Hence,  Eq. (\ref{normxi}) with $\rho  =  I/2$  implies
  $  \left( \frac    {4a +   2 b_{11} }  3 \right)   P_1  +  2b_{00} P_0  =    \frac  {I \otimes I}{2} $,
which is equivalent to the relations 
 $2a +  b_{11}  =  \frac 3 4$ and $b_{00}  =  \frac 14 $.   
On the other hand, direct calculation shows that the overlap $\Tr[\Gamma_0\Gamma_1]$ is zero if and only if $a=b_{11} =  b_{01}  = 0$   (see the Appendix).   Since this condition is incompatible with the normalization condition $2a +  b_{11}  =  \frac 3 4$, we proved that no circuit with  causal structure $A \preceq B$ can perfectly discriminate  between $\map C_0$ and $\map C_1$ with a single query.    Moreover, since the Choi operators $C_0$ and $C_1$ are invariant under the exchange $(A, A')  \leftrightarrow (B,B')$, the same derivation can be used to prove that perfect discrimination cannot be achieved with a single query by any circuit with  causal structure $B\preceq A$. In conclusion, perfect discrimination in a circuit with fixed causal structure requires at least two queries.   This number is actually sufficient, because with two queries the quantum superposition of causal structures can be simulated in an ordinary circuit  using controlled swap operations \cite{qswitch}.  

Iterating the result for $N$ different pairs of no-signalling channels $\{\map C^{(n)}_{i_n}~|~  i_n  \in \{0,1\},  n= 1, \dots N \}$  it is easy to see that  superposing two causal  structures for each pair allows us to distinguish with probability 1 among $2^N$ channels using one query to the black box $\bigotimes_{n=1}^N   \map C_{i_n}^{(n)}$.  Without the superposition of causal structures, the probability of successfully distinguishing all pairs of channels with a single query would go to zero exponentially fast in $N$.  This fact can be the product rule of Ref.  \cite{inprep} shows that the maximum probability  $p^{(N)}_{succ}$ of distinguishing correctly all  the $N$ channels is equal to the product of the probabilities of distinguishing each channel separately, that is, $p^{(N)}_{succ}  =  [p^{(1)}_{succ}]^{N}   \to 0$.

Before concluding,  it is worth highlighting a remarkable feature of our result:  Perfect discrimination is achieved by superposing two strategies that, considered separately, are very inefficient.  It is easy to show that the probability of success of the strategies  a) and  b) in Eq. (\ref{duecircuiti}) is $p_{succ}^{(a)}  =  p_{succ}^{ (b)}  =  \frac 2 3$, a value that can be easily beaten even by parallel strategies. For example,  applying the unknown channel   $\map C_i$, $i= 0,1$ on one side of a maximally entangled state, thus obtaining the Choi state $\rho^{Choi}_i  =  C_i  /4$, $i=0,1$, yields the much higher success  probability $p_{succ}^{Choi} =   \frac {11} {12} $.  
Quite paradoxically, it is exactly by superposing two sub-optimal strategies
that one can achieve perfect discrimination. This feature suggests an analogy
with Parrondo's paradox in classical game theory \cite{parrondo}, where the
alternate choice of two losing games yields a winning game (i.e. a
game where the optimal strategy yields a winning probability larger
than 1/2).  In the quantum example the counterintuitive feature is
even more striking: The probability of winning the discrimination game
jumps to $p_{succ} =1$ thank to the quantum
superposition of the losing strategies a) and b).

In conclusion,  we demonstrated that  the quantum superposition of causal structures is a new primitive that offers an advantage over causally ordered quantum circuits  in the  problem of quantum channel discrimination.  
Such a result is similar in spirit to that of Oreshkov, Costa, and Brukner \cite{costa}, who showed the advantage of non-causal strategies in a non-local (Bell-inequality-type) game.     
These  results, along with the quantum switch of Ref. \cite{qswitch},  are starting to unveil some deep relationship between quantum theory, causal order and space-time, and more developments in this direction are expected to come in the near future.


\emph{Acknowledgments.}   I acknowledge G. M. D'Ariano, P. Perinotti, J. Watrous,  and M. Piani for stimulating discussions and two anonymous Referees for useful comments.  This work is supported  the National Basic Research Program of China (973) 2011CBA00300 (2011CBA00301).

\appendix  
\begin{widetext}

\section{Expanded proof}

The following appendices contain  an expanded version of the proof that no circuit with fixed causal structure can perfectly distinguish between the channels C0 and C1 with a single query to the black box.
The proof that no quantum circuit with fixed causal structure can distinguish between $\map C_0$ and $\map C_1$ with a single query consists of the following steps:   
\begin{enumerate}
\item Show that without loss of generality we can restrict to symmetric testers $\{T_0, T_1\}$, satisfying $[T_i,  U\otimes U^* \otimes U \otimes U^*]  = 0, \forall U\in SU (2)$
\item Show that the normalization of the tester $\{T_0, T_1\}$ is equivalent to the relations  $2a +  b_{11}  =  \frac 3 4$ and   $  
b_{00}  =  \frac 14 $, where $a$ and $(b_{ij})_{i,j=0,1}$ are the coefficient in the expression  $
\widetilde \Xi  =  a  P_{\frac 3 2}  + \sum_{m,n=0,1}   b_{mn}  ~  T_{\frac 1 2}^{mn}$  [we recall that $\widetilde \Xi$ is the operator on $\spc H_3 \otimes \spc H_2 \otimes \spc H_1$ defined as $\widetilde \Xi  :  =  (Y  \otimes I_2  \otimes  Y)  \Xi  (Y  \otimes I_2  \otimes  Y) $, with $\Xi$ defined by the relation $T_0  +  T_1  =  I_4 \otimes \Xi$]   
\item  Prove that that perfect discrimination between the channels $\map C_0$ and $\map C_1$ is equivalent to perfect discrimination between the states $\Gamma_0$ and $\Gamma_1$, defined by $\Gamma_i  : =  (I_4 \otimes \Xi^{\frac 12}) C_i  (I_4 \otimes \Xi^{\frac 12})$, and, therefore, is equivalent to   the zero-overlap condition 
\begin{align}
0 =  \Tr[\Gamma_0\Gamma_1]  =  \Tr \left[   \widetilde C_0 \left(I_4 \otimes \widetilde \Xi    \right)   \widetilde C_1   \left(  I_4 \otimes \widetilde \Xi   \right)  \right],
\end{align}  
with $\widetilde C_i, ~i = 0,1$ defined as  $\widetilde C_i  :  =  (Y  \otimes I_2  \otimes  Y)  C_i  (Y  \otimes I_2  \otimes  Y) $, $C_i$ being the Choi operator of channel $\map C_i$.   
\item Compute  the overlap  $\Tr[\Gamma_0\Gamma_1]$ and show that it is zero if and only if $a =  b_{11}  =  b_{01}  = 0$
\item Observe that the zero overlap condition $a =  b_{11}  =  b_{01}  = 0$ is incompatible with the condition $2a +  b_{11}  =  \frac 3 4 $ in the normalization of the tester  $\{T_0, T_1\}$.     
\end{enumerate}

Steps 1,  2 and 3 have been already addressed  in the main text, while step 5 is  obvious from steps 2  and 4.    The only point that we still need to address is step 4, which requires the calculation of the overlap $\Tr[\Gamma_0\Gamma_1]$. This will be done in the next section. 
\section{Calculation of the overlap between $\Gamma_0$ and $\Gamma_1$}

Since the overlap is given by $ \Tr[\Gamma_0\Gamma_1]  =  \Tr \left[   \widetilde C_0 \left(I_4 \otimes \widetilde \Xi    \right)   \widetilde C_1   \left(  I_4 \otimes \widetilde \Xi   \right)  \right]$ to compute it we first need the explicit expressions for $\widetilde C_0, \widetilde C_1$ and $  I_4 \otimes \widetilde \Xi$.  They will be worked out in the next subsections \ref{subsec:cc} and \ref{subsec:xi}.

\subsection{Expressions for $\widetilde C_0$ and $\widetilde C_1$}\label{subsec:cc}

By definition of $\widetilde C_i$,  we have  $\widetilde C_i  :  =  (Y  \otimes I_2  \otimes  Y)  C_i  (Y  \otimes I_2  \otimes  Y) $, $C_i$ being the Choi operator of channel $\map C_i$.   Now, the Choi operator $C_i$ is given by Eq. (10)  of the main text, which reads  
\begin{align}\label{avechannel}
C_i   : =  \int \d U  ~  (\map U  \otimes \map U^*  \otimes \map U \otimes \map U^*) ( \Lambda_i )  
\end{align}
where 
\begin{align}
\Lambda_0  &=   \sum_{m,n = 0,1}    |m\>\<m|  \otimes   |m\>\<m|  \otimes  |n\>\<n|  \otimes  |n\>\<n| ,  \\ 
\Lambda_1&= |Y\>\!\>  \<\!\<  Y|  \otimes   |X\>\!\> \<\!\<  X| .
\end{align}

Using  the relations $YU^*  Y  =  U$ and $Y^2  = I$ we then obtain
        \begin{equation}\label{brrr}
    \widetilde C_i    =   \int \d U  ~     \map U^{\otimes 4}  (\widetilde \Lambda_i),
    \end{equation} 
    with  
\begin{align}
\widetilde \Lambda_{0}  & :=     (I \otimes  Y  \otimes I \otimes  Y )  \Lambda_0 (I \otimes  Y \otimes I \otimes  Y )  =   \sum_{m, n  = 0,1  } |m\>\<  m|  \otimes   |  m  \oplus 1 \>\<  m \oplus  1| \otimes  |n\>\<  n| \otimes  |n \oplus 1\>\<  n \oplus 1|       \\
\widetilde \Lambda_{1} & :=        (I \otimes  Y  \otimes I \otimes  Y )  \Lambda_0  (I \otimes  Y \otimes I \otimes  Y )   =  |I\>\!\> \<\!\<  I|   \otimes |Z\>\!\>\<\!\<  Z| ,
\end{align}
$\oplus$ denoting here the addition modulo 2.

To do the integral in Eq. (\ref{brrr}), we need to find the components of $\widetilde \Lambda_1$ on the invariant subspaces of the tensor representation $U^{\otimes 4}$.   For convenience of reading, when writing vectors  we will order the Hilbert spaces as $\spc H_1 \otimes \spc H_2  \otimes \spc H_3 \otimes \spc H_4$  instead of    $\spc H_4 \otimes \spc H_3  \otimes \spc H_2 \otimes \spc H_1$.   
We have 
\begin{align}
\nonumber |Z\>\!\>_{1,2} \otimes |I\>\!\>_{3,4}  &  =  |1111\>  -   |0000\>  +  |1100\>  -  |0011\>  \\
  &  = |2,2 ; (1,1)\>  -  |2,-2; (1,1) \>  +  \sqrt 2   |1,0;(1,1)\>,      
\end{align} 
where $|j,m;(k,l)\>$ is the eigenstate of the $z$-component of the angular momentum with eigenvector $m$, in the subspace with total angular momentum $j$ obtained by the tensor product of two subspaces where spins 1 and 2 have total angular momentum $k$ and spins 3 and 4 have total angular momentum $l$.
Taking the average of $\widetilde \Lambda_1$ we then obtain 
\begin{align}
\widetilde C_1  =  2 \frac {P_{2;(1,1)}}{d_2 }   +  2  \frac {P_{1;(1,1)}}{d_1},
\end{align}
where $d_j  =  2j +1$. 
To find the components of $\widetilde \Lambda_0$ on the invariant subspaces we express it as 
\begin{align}
\widetilde \Lambda_0  =  (   |1,0  \>  \<  1,0 |  +  |0,0\>\<0,0|  )_{1,2}  \otimes    (   |1,0  \>  \<  1,0 |  +  |0,0\>\<0,0|  )_{3,4}    
\end{align}
where $|1,0\>  :  =  \frac{ |10\>  +  |01\>}{\sqrt 2}  $ and $|0,0\>  :  =  \frac{ |10\>  -  |01\>}{\sqrt 2}  $.  Now, the vector $|1,0\>_{1,2}|1,0\>_{3,4} $ can be decomposed as 
\begin{align}
|1,0\>_{1,2}|1,0\>_{3,4}   =  \sqrt{\frac 23} |2, 0; (1,1)\>   -  \sqrt{\frac 13}  |0,0;  (1,1)\>,  \end{align}
so that we have  
\begin{align}
\nonumber \widetilde \Lambda_0    =  & \left  (  \sqrt{\frac 23} |2, 0; (1,1)\>   -  \sqrt{\frac 13}  |0,0;  (1,1)\> \right)   \left(  \sqrt{\frac 23} \< 2, 0; (1,1)|   -  \sqrt{\frac 13}  \<0,0;  (1,1)|  \right)  +\\
  &  +  |1,0;(1,0)  \>     \<1, 0;(1,0)  |   +    |1,0;(0,1)  \>     \<1, 0;(0,1)  |   +  |0,0;(0,0)  \>     \<0, 0;(0,0)  | .
\end{align}  
Hence, taking the average of $\widetilde \Lambda_0$ we obtain  
\begin{align}
\widetilde C_0  =    \frac 23 \frac {P_{2; (1,1)}}{d_2  }   +  \frac 13 P_{0;(1,1)}  +  \frac{P_{1; (1,0) }  + P_{1; (0,1) }    } {d_1} +    P_{0; (0,0)}  
\end{align}

\subsection{Expression for $I_4 \otimes \widetilde \Xi$}\label{subsec:xi}
Recall that, due to the symmetry of the tester,  the operator $\widetilde \Xi$  has the expression  
\begin{align}\label{encore}
\widetilde \Xi  =  a  P_{\frac 3 2}  + \sum_{m,n=0,1}   b_{mn}  ~  T_{\frac 1 2}^{mn}.    
\end{align}
[cf. Eq. (13)  of the main text].
By Eq. (\ref{encore}),  we have  $\widetilde \Xi  =  a P_{\frac 32}  +    K_{\frac 12}  $, where $K_{\frac 12} : =  \sum_{m,n=0,1}  b_{mn}  T_{\frac 12}^{mn}$ is a positive operator with support contained in an invariant (although not necessarily irreducible) subspace with total angular momentum $j=  \frac 12$.   Hence, we will have  
\begin{align}\label{tensor}
I_4 \otimes \widetilde \Xi  =   a  (  Q_{2; (\frac 32,\frac 12)}   +    Q_{1; (\frac 32,\frac 12)}  )    +  L_{1;(\frac 12 , \frac 12)}   +     L_{0;(\frac 12 , \frac 12)},
\end{align}
where $Q_{j; (k,l)}$ is the projector on the subspace with total angular momentum $j$, resulting from the tensor product of the Hilbert space $\spc H_4$ with the subspace of $\spc H_1 \otimes \spc H_2 \otimes \spc H_3$   with total angular momentum $k$  and $L_{j; (\frac 12,\frac 12)}$ is a positive operator with support contained in the subspace with total angular momentum $j$, resulting from the tensor product of the Hilbert space $\spc H_4$ with the subspace of $\spc H_1 \otimes \spc H_2 \otimes \spc H_3$   with total angular momentum $\frac 12$.   Note that, since the representation with $j=2$ has unit multiplicity, we necessarily have $Q_{2,:  (\frac 32 ,\frac 12 )}  \equiv  P_{2:  (1,1)}$, where we recall that $P_{j:k,l}$ was defined as the projector on the subspace with total angular momentum number $j$ resulting from the tensor product of the two subspaces of $\spc H_4 \otimes \spc H_3$  and $\spc H_2 \otimes \spc H_1$ with total angular momenta $k$ and $l$, respectively.     

\subsection{The zero-overlap condition}
Here we calculate the overlap $\Tr[\Gamma_0 \Gamma_1]$ and show that it vanishes if and only if the coefficients $a, b_{11}$ and $b_{01}$ vanish.  
To start the calculation, recall the expression of the overlap between $\Gamma_0$ and $\Gamma_1$, given by  
\begin{align}\label{encore2}
\Tr[\Gamma_0  \Gamma_1]  =  \Tr \left[   \widetilde C_0 \left(I_4 \otimes \widetilde \Xi    \right)   \widetilde C_1   \left(  I_4 \otimes \widetilde \Xi   \right)  \right]
\end{align} 

Inserting the expressions for $\widetilde \Xi$, $\widetilde C_0$ and $\widetilde C_1$  in Eq. (\ref{encore2}), we obtain:
\begin{align}\label{sum}
\nonumber \Tr[\Gamma_0 \Gamma_1]  = &  \frac {4}{3d^2_2}    \Tr [  P_{2; (1,1)}    (I_4 \otimes \widetilde \Xi    )    P_{2; (1,1)}  (I_4  \otimes \widetilde \Xi ) ]    +       \frac {4 }{3 d_2 d_1 }    \Tr [  P_{2; (1,1)}    (I_4 \otimes \widetilde \Xi    )    P_{1; (1,1)}  (I_4  \otimes \widetilde \Xi ) ] +     \\  
\nonumber  &  +   \frac {2}{3d_2}    \Tr [  P_{0; (1,1)}    (I_4 \otimes \widetilde \Xi    )    P_{2; (1,1)}  (I_4  \otimes \widetilde \Xi ) ]   +       \frac {2 }{3  d_1 }    \Tr [  P_{0; (1,1)}    (I_4 \otimes \widetilde \Xi    )    P_{1; (1,1)}  (I_4  \otimes \widetilde \Xi ) ]  +\\  
\nonumber &  +  \frac {2 }{ d_1 d_2}    \Tr [ ( P_{1; (1,0)}    +   P_{1; (0,1)} )    (I_4 \otimes \widetilde \Xi    )    P_{2; (1,1)}  (I_4  \otimes \widetilde \Xi ) ]  +   \frac {2 }{ d_1^2 }    \Tr [ ( P_{1; (1,0)}    +   P_{1; (0,1)} )    (I_4 \otimes \widetilde \Xi    )    P_{1; (1,1)}  (I_4  \otimes \widetilde \Xi ) ] + \\
  \nonumber & +   \frac {2 }{  d_2}    \Tr [  P_{0; (0,0)}        (I_4 \otimes \widetilde \Xi    )    P_{2; (1,1)}  (I_4  \otimes \widetilde \Xi ) ]    +   \frac {2 }{  d_1}    \Tr [  P_{0; (0,0)}        (I_4 \otimes \widetilde \Xi    )    P_{1; (1,1)}  (I_4  \otimes \widetilde \Xi ) ] \\
\nonumber =   &  \frac {4 a^2}{3d^2_2}    \Tr [  P_{2; (1,1)}     ]    +        0  +  \\
\nonumber   & +  0  + 0 + \\
\nonumber &  + 0  +   \frac {2 }{ d_1^2 }    \Tr [ ( P_{1; (1,0)}    +   P_{1; (0,1)} )    (I_4 \otimes \widetilde \Xi    )    P_{1; (1,1)}  (I_4  \otimes \widetilde \Xi ) ] +  \\
 \nonumber  & +  0     +   0\\
   =&   \frac {4 a^2}{3d_2}   +    \frac {2 }{ d_1^2 }    \Tr [   P_{1; (1,0)}    (I_4  \otimes \widetilde \Xi )      P_{1; (1,1)}      (I_4  \otimes \widetilde \Xi )    ]  +   \frac {2 }{ d_1^2 }    \Tr [  P_{1; (0,1)}       (I_4 \otimes \widetilde \Xi    )    P_{1; (1,1)}  (I_4  \otimes \widetilde \Xi ) ] 
\end{align}
Now, the three terms in the sum are all non-negative.  Hence, in order to have $\Tr [\Gamma_0 \Gamma_1]  =  0$ they must all vanish.  In particular, we must have   $a= 0$, whence Eq. (\ref{tensor})   becomes
\begin{align}\label{mamma}
I_4 \otimes \widetilde \Xi  =    L_{1;(\frac 12 , \frac 12)}   +     L_{0;(\frac 12 , \frac 12)},
\end{align}  
and the  overlap in Eq. (\ref{sum})  becomes
\begin{align}
\label{suma}
 \Tr[\Gamma_0 \Gamma_1]  =          \frac {2 }{ d_1^2 }    \Tr [   P_{1; (1,0)}    L_{1; (\frac 12 , \frac 12)}   P_{1; (1,1)}     L_{1; (\frac 12 , \frac 12)}   ]  +   \frac {2 }{ d_1^2 }    \Tr [  P_{1; (0,1)}     L_{1; (\frac 12 , \frac 12)}     P_{1; (1,1)}   L_{1; (\frac 12 , \frac 12)}  ] 
\end{align}
To continue the calculation we  now need to find the explicit expression for $ L_{1; (\frac 12 , \frac 12)} $. To find it, we first decompose the tensor product $\spc H_4 \otimes \spc L_{\frac 12}^{(m)}$, with $m= 0,1$ as $  \spc H_4 \otimes \spc L_{\frac 12}^{m}   =  \spc K_1^{(m)}  \oplus    \spc K^{(m)}_0$, where  the irreducible subspaces $\spc K_0^{(m)}$ and  $\spc K_1^{(m) }$ have total angular momentum $j=0$ and $j=1$, respectively.  
In particular, we are interested in the $j=1$ subspaces:   $ \spc K_1^{(0)}$  is spanned by the vectors 
\begin{align*}
|\Psi_{1,1}^{(0)}\>  &:  =   \frac{ |Y\>\!\>_{1,2}  |1\>_3  |1\>_4 }{\sqrt 2}  \\
|\Psi_{1,0}^{(0)}\>  &:  =   \frac{ |Y\>\!\>_{1,2} |X\>\!\>_{34}}{2}\\ 
|\Psi_{1,-1}^{(0)}\>  &:  =   \frac{ |Y\>\!\>_{1,2}  |0\>_3  |0\>_4 }{\sqrt 2}    ,
\end{align*} 
while $ \spc K_1^{(1)}$  is spanned by the vectors 
\begin{align*}
|\Psi_{1,1}^{(1)}\>    &:  =   \frac{  |1\>_1  |Y\>\!\>_{2,3}     |1\>_4  + |Y\>\!\>_{1,3} |1\>_2  |1\>_4  }{\sqrt 6}  \\
|\Psi_{1,0}^{(1)}\>  :&    =   \frac{  |X\>\!\>_{14}  |Y\>\!\>_{2,3}       + |Y\>\!\>_{1,3} |X\>\!\>_{2,4}  }{2 \sqrt {3}}   \\ 
|\Psi_{1,-1}^{(1)}\>  &:  =       \frac{  |0\>_1  |Y\>\!\>_{2,3}     |0\>_4  + |Y\>\!\>_{1,3} |0\>_2  |0\>_4  }{\sqrt 6}     ,
\end{align*} 
Comparing the two sides of Eq. (\ref{mamma}) we obtain  $ L_{1; (\frac 12 , \frac 12)}   =  \sum_{m,n=0,1}   b_{mn}   S^{mn}_{1}$, with $S^{mn}_1  :  =  \sum_{k=  -1}^1  |\Psi_{1,k}^{(m)}\>  \<  \Psi_{1,k}^{(n)}|   $.  
Evaluating the right-hand-side of Eq. (\ref{suma}) we get  
\begin{equation}\label{cacca}
\Tr[\Gamma_0 \Gamma_1]  =          \frac {2 |b_{01}|^2  }{ d_1^2 }    \Tr [   P_{1; (1,0)}  S^{01}_1  P_{1; (1,1)}    T^{10}_1    ]  +   \frac {2 b_{11}^2 }{ d_1^2 }    \Tr [  P_{1; (0,1)}   S^{11}_1     P_{1; (1,1)}   T^{11}_1  ] .
\end{equation}
Now, inserting in the above expression the definition $S^{mn}_1 :  =  \sum_{k=  -1}^1  |\Psi_{1,k}^{(m)}\>  \<  \Psi_{1,k}^{(n)}|  $ for $m,n =0,1$,   
   we obtain  
\begin{align}
 \nonumber \Tr [   P_{1; (1,0) }  S^{01}_1  P_{1; (1,1)}    S^{10}_1    ]   &= \sum_{k,l=-1}^1    \<  \Psi_{1,l}^{(0)}|        P_{1; (1,0)}      |\Psi_{1,k}^{(0)}\>  
   \<  \Psi_{1,k}^{(1)}|    P_{1; (1,1)}    |\Psi_{1,l}^{(1)}\>  \\ 
  \label{cacca1}  &= \sum_{k=-1}^1    \<  \Psi_{1,k}^{(0)}|        P_{1; (1,0)}      |\Psi_{1,k}^{(0)}\>  
   \<  \Psi_{1,k}^{(1)}|    P_{1; (1,1)}    |\Psi_{1,k}^{(1)}\>   
 \end{align}
and 
\begin{align}
\nonumber  \Tr [   P_{1; (0,1)}  S^{11}_1  P_{1; (1,1)}    T^{11}_1    ]    &   = \sum_{k,l=-1}^1    \<  \Psi_{1,l}^{(1)}|        P_{1; (0,1)}      |\Psi_{1,k}^{(1)}\>           \<  \Psi_{1,k}^{(1)}|    P_{1; (1,1)}    |\Psi_{1,l}^{(1)}\>  
\\ 
\label{cacca2}  &  = \sum_{k=-1}^1    \<  \Psi_{1,k}^{(1)}|        P_{1; (0,1)}      |\Psi_{1,k}^{(1)}\>           \<  \Psi_{1,k}^{(1)}|    P_{1; (1,1)}    |\Psi_{1,k}^{(1)}\>  \end{align}
To conclude the calculation, we express the projectors     $P_{1;(1,0)}$,   $P_{1; (0,1)}  $,  and    $P_{1; (1,1)}$ as  
\begin{align}   
 P_{1; (1,0)}   & =  \sum_{k=-1}^1  |V_k  \>\<V_k|    \\ 
   P_{1; (0,1)}    &  =  \sum_{k=-1}^1  |W_k  \>\<W_k|  \\
      P_{1; (1,1)}  &  =  \sum_{k=-1}^1  |Z_k  \>\<Z_k|  ,
      \end{align} 
 with
\begin{align*} 
|V_1  \>  &:  =    \frac{ |Y\>\!\>_{1,2}  |1\>_3  |1\>_4 }{\sqrt 2}  \equiv     |\Psi_{1,1}^{(0)}\>  \\ 
   |V_0\> &:  =  \frac{ |Y\>\!\>_{1,2} |X\>\!\>_{3,4}}{2} \equiv |\Psi_{1,0}^{(0)}\>   \\ 
   |V_{-1}\>  &:  =    \frac{ |Y\>\!\>_{1,2}  |0\>_3  |0\>_4 }{\sqrt 2}   \equiv |\Psi_{1,-1}^{(0)}\>   \\ 
 |W_1  \>  &:  =      \frac{    |1\>_1 |1\>_2 |Y\>\!\>_{3,4} }{\sqrt 2}    \\
   |W_0\> &:  =        \frac{ |  X\>\!\>_{12}  |Y\>\!\>_{3,4} }{2}     \\    
   |W_{-1}\>  &:  =     \frac{   |0\>_1  |0\>_2   |Y\>\!\>_{3,4}  }{\sqrt 2}     \\
   |Z_1\>  &:  =  \frac{  |1\>_1 |1\>_2  |X\>\!\>_{3,4}  -  |  X\>\!\>_{1,2}   |1\>_3  |1\>_4  }{2}  \\
    |Z_0\>&:  =  \frac{  |1\>_1 |1\>_2  |0\>_3  |0\>_4   -  |0\>_1|0\>_2   |1\>_3  |1\>_4  }{\sqrt 2}  \\
     |Z_{-1}\>&:  =  \frac{   |  X\>\!\>_{1,2}   |0\>_3 |0\>_4   -   |0\>_1 |0\>2  |X\>\!\>_{3,4} }{ 2}  
\end{align*}
Computing the overlaps 
\begin{align*}
\< \Psi^{(0)}_{1,k}  |V_k\>  &=  1    \qquad    \forall k =-1,0,1\\ 
\< \Psi^{(1)}_{1,k}  |W_k\>  &=  \frac {-1}{\sqrt 3}  \qquad  \forall k =-1,0,1\\ 
\< \Psi^{(1)}_{1,k}  |Z_k\>  &=  \sqrt{\frac 23}  \qquad   \forall k =-1,0,1\\ 
\end{align*}
and inserting them in  Eqs. (\ref{cacca1}), (\ref{cacca2})  we finally obtain 
\begin{align*}
 \Tr [   P_{1; (1,0)}  S^{01}_1  P_{1; (1,1)}    S^{10}_1    ]    &=  \sum_{k=-1}^1     |   \<  \Psi_{1,k}^{(0)}|      V_k\>   |^2    
  | \<  \Psi_{1,k}^{(1)}|  Z_k\> |^2  =2        \\
  \Tr [   P_{1; (0,1)}  S^{11}_1  P_{1; (1,1)}    S^{11}_1    ]   &  =  \sum_{k=-1}^1     | \<  \Psi_{1,k}^{(1)}|  W_k \> |^2          |   \<  \Psi_{1,k}^{(1)}|      Z_k\> |^2      =  \frac 23 \\
\end{align*}
so that Eq. (\ref{cacca}) becomes   
\begin{equation}  
\Tr[\Gamma_0 \Gamma_1]  =          \frac {4 |b_{01}|^2  }{ d_1^2 }   +   \frac {4 b_{11}^2 }{3 d_1^2 }  .
\end{equation}
In conclusion, we showed that the condition $\Tr[\Gamma_0 \Gamma_1] =  0$ holds if and only if $a  =  b_{11}  =  b_{01}  =  0$.  As already anticipated,  this condition is in contradiction with the normalization of the tester $\{T_0,T_1\}$, which imposes  $2a +  b_{11}  =  \frac 3 4 $.   
In conclusion, this  proves that there cannot exist any sequential strategy that  can distinguish perfectly between $\map C_0$ and $\map C_1$  using a single query to the black box.  
 
\end{widetext}
 

\begin{thebibliography}{55}
\bibitem{purificaz}  The restriction to pure states and isometries can be done without loss of generality thanks to the purification property of quantum theory.
\bibitem{nosig}  Note that no-signalling is defined as an information-theoretic property of the black box.  Here we do not assume that Alice and Bob are spacelike separated.   What we assume instead is that the no-signalling box is able to provide Alice with an output as soon as she  supplies an input, independently of  Bob's actions, and vice-versa. This assumption is very natural in the case of product channels, of the form $\map A \otimes \map B$, and of randomizations of product channels, of the form  $\sum_k  p_k  \map A_k \otimes \map B_k$ for some probabilities $\{p_k\}$.
\end{thebibliography}
\end{document}